\documentclass[prd,aps,floatfix,nofootinbib,preprint,showpacs,tightenlines]{revtex4}
\usepackage{dcolumn}% Align table columns on decimal point
\usepackage{amsmath}
\usepackage{latexsym}
\usepackage{graphicx}
\usepackage{bm}

\def\svev#1{\left\langle #1\right\rangle}       % variable < >
\def\tr{{\rm tr}\,}
\def\Tr{{\rm Tr}\,}

\def\Re{{\rm Re\,}}

\long \def \blockcomment #1\endcomment{}
\def\det{{\rm det}}
\def\Det{{\rm Det}}
\def\Seff{S_\textrm{eff}}
\def\Veff{V_{\textrm{eff}}}
\def\Vf{V_f}

\def\Eq#1{Eq.~(\ref{#1})}
\def\bx{\mathbf{x}}
\def\Li{\textrm{Li}}
\def\ZN{$\textrm{Z}(N)$}

\begin{document}
%%%%%%%%%%%%%%%%%%%%%%%%%%%%%%%%%%%%%%%%%%%%%%%%%%%%%%%%%%%%%%%%%%%%%%
\title{%
Metastable nonconfining states in SU(3) lattice gauge theory with sextet fermions}
\author{Olga Machtey}
\author{Benjamin Svetitsky}
\affiliation{Raymond and Beverly Sackler School of Physics and Astronomy, Tel~Aviv University, 69978 Tel~Aviv, Israel}

\begin{abstract}
We study the SU(3) lattice gauge theory, with  two flavors of sextet Wilson-clover fermions, near its finite-temperature phase transition.
We find metastable states that have Wilson line  expectation values whose complex phases are near $\pm2\pi/3$ or $\pi$.  The true equilibrium phase at these couplings and temperatures has its Wilson line oriented only towards the positive real axis, in agreement with perturbation theory.
\end{abstract}

\maketitle

%%%%%%%%%%%%%%%%%%%%%%%%%%%%%%%%%%%%%%%%%%%%%%%%%%%%%%%%%%%%%%%%%%%%%
\section{Introduction}
%%%%%%%%%%%%%%%%%%%%%%%%%%%%%%%%%%%%%%%%%%%%%%%%%%%%%%%%%%%%%%%%%%%%%
SU($N$) gauge theories with fermions in two-index color representations have attracted attention recently for several reasons.
One is their role in technicolor theories that might be viable as extensions of the Standard Model \cite{Sannino:2004qp,Dietrich:2006cm}.
Another is the appearance of matter fields in two-index representations as components of orientifold equivalences \cite{Armoni:2003gp,Unsal:2006pj,Armoni:2007rf,Armoni:2007kd}.
In connection with the first,
we have been studying the lattice SU(3) gauge theory with fermions in the two-index symmetric representation, which is the sextet \cite{Shamir:2008pb,DeGrand:2008kx}.
In the course of our study of its phase diagram we found high-temperature states that appear to possess a Wilson line condensate 
$\svev L$ that lies along one of several directions in the complex plane, with apparent phase transitions among these non-confining states (see also \cite{DeGrand:2009et, DeGrand:2009hu,Fodor:2009wk,Sinclair:2009ec}).
Condensates of this nature are known to occur when the fermions are made to obey periodic boundary conditions \cite{Unsal:2006pj,DeGrand:2006qb,DeGrand:2007tw,Myers:2008ey,Myers:2009df}, but not when the boundary conditions are antiperiodic as appropriate for the time direction in the finite-temperature theory~\cite{Gocksch:1993iy}.
Thus these states merited further study.

A strong-coupling calculation, applicable to the low-temperature disordered phase, shows that sextet fermions tend to deflect the disordered Wilson line towards the negative real axis.
If this were to persist into intermediate couplings and across the finite-temperature phase transition, we would obtain a condensate with a negative real part.
Weak-coupling analysis, on the other hand, gives the effective potential for $L$ a unique minimum
with phase $\theta=0$; nonetheless, there are local minima at $\theta=\pm2\pi/3$ and at $\theta=\pi$.
Hence one could conceive of a global minimum appearing in one of these other directions as the coupling is made stronger.

In the following, we first review the analytical weak- and strong-coupling approximations to the effective potential.
Turning to numerical work, we show the results of simulating the
SU(3) gauge theory with two flavors of Wilson-clover fermions, in the sextet representation, at intermediate couplings on a finite-temperature lattice.
Quenching a disordered configuration into the high-temperature regime indeed shows that the Wilson line can condense into ordered states in various directions in the complex plane, depending on the gauge coupling $\beta$ and hopping parameter $\kappa$.
The phase angle of the condensate can take values near
$\theta=0$, $\pm2\pi/3$, and $\pi$ (the last comes of a partial disordering of the $\theta=\pm2\pi/3$ states).
We even find tunneling and coexistence among these states.
In order to settle which is the true equilibrium state, we simulate at these values of $(\beta,\kappa)$ from mixed initial conditions, putting half the initial lattice into an ordered configuration
with $\theta=0$.
We find that in all cases, the Wilson line condensate comes to equilibrium in the direction $\theta=0$.
Thus the other orientations of the condensate have higher free energy and are merely metastable.

%%%%%%%%%%%%%%%%%%%%%%%%%%%%%%%%%%%%%%%%%%%%%%%%%%%%%%%%%%%%%%%%%%%%%
\section{Definitions}
%%%%%%%%%%%%%%%%%%%%%%%%%%%%%%%%%%%%%%%%%%%%%%%%%%%%%%%%%%%%%%%%%%%%%
Let us briefly review the symmetries of the pure gauge theory, their action on the Wilson line, and how they are broken by matter fields
\cite{Svetitsky:1985ye}.
We discuss the continuum SU($N$) gauge theory at finite temperature
$\beta^{-1}$.

The Wilson line is defined as
\begin{equation}\label{eq:Wilson-line}
		L(\mathbf{x})=\frac{1}{N}\,\textrm{tr\,}\Omega(\mathbf{x}),
\end{equation}
where $\Omega(\mathbf{x})$ is the path-ordered exponential,
\begin{equation}
\Omega(\mathbf{x})=P\exp\left[i\int_0^\beta A_{0}(\mathbf{x},\tau)\,d\tau\right].
\label{Omega}
\end{equation}
Here $A_{0}(\mathbf{x},\tau)=A_{0}^{a}(\mathbf{x},\tau)\lambda^{a}$
is the gauge field in the fundamental representation; the gauge
fields $A_{\mu}^{a}(\mathbf{x},\tau)$ satisfy periodic
boundary conditions in the Euclidean time direction, $A_{\mu}^{a}(\mathbf{x},0)=A_{\mu}^{a}(\mathbf{x},\beta)$.

The pure gauge theory is invariant under gauge transformations that are periodic up to a \ZN~phase $z$,
\begin{equation}
\label{eq:gauuge_trans}
	U(\mathbf{x},0)=zU(\mathbf{x},\beta),
\end{equation}
where $z^N=1$.
If we denote by $\cal G$ the group of strictly periodic gauge transformations, the complete symmetry group of the pure gauge theory
is ${\cal G}\times\textrm{Z}(N)$.
The Wilson line, invariant under ${\cal G}$, transforms as
\begin{equation}
L(\bx)\to zL(\bx).
\end{equation}
In the low-temperature phase of the pure gauge theory, the global \ZN~symmetry is unbroken and hence $\svev L=0$.
In the high-temperature phase the symmetry is spontaneously broken, permitting $\svev L =L_0\not=0$; there are $N$ equivalent equilibrium states, characterized by $\svev L=z L_0$.

A fermion field in representation $R$ will transform as\begin{equation}
		\psi\left(\mathbf{x},\tau\right)\rightarrow U_{R}\left(\mathbf{x},\tau\right)\psi\left(\mathbf{x},\tau\right).
\end{equation}
If $U$ is aperiodic as in \Eq{eq:gauuge_trans},
then $U_R$ will be aperiodic according to
\begin{equation}
	U_R(\mathbf{x},0)=z^jU_R(\mathbf{x},\beta),
\end{equation}
where $j$ is the $N$-ality of $R$.
This transformation will not generally preserve the anti-periodic boundary conditions imposed on the fermions (unless $j=0$).
Thus the fermion action breaks the \ZN~symmetry explicitly.

%%%%%%%%%%%%%%%%%%%%%%%%%%%%%%%%%%%%%%%%%%%%%%%%%%%%%%%%%%%%%%%%%%%%%
\section{Weak coupling}
%%%%%%%%%%%%%%%%%%%%%%%%%%%%%%%%%%%%%%%%%%%%%%%%%%%%%%%%%%%%%%%%%%%%%
The one-loop calculation of the effective potential is essentially a continuum calculation.
It was carried out long ago \cite{Gross:1980br,Weiss:1980rj} for massless fermions; for massive fermions, which concern us here, see \cite{Myers:2009df}.
One writes
\begin{equation}
		A_{\mu}=a_{\mu}+g\delta A_{\mu},
\end{equation}
where $a_{\mu}=(\Phi/\beta)\delta_{\mu0}$ is a constant background field that gives a background Wilson line via $\Omega=e^{i\Phi}$.
We may choose a gauge where $\Phi$, and hence $\Omega$, is diagonal.
Performing the gaussian integration over $\delta A_{\mu}$ in background gauge, one obtains the effective potential,
\begin{equation}
	e^{-\beta \Veff\left[\Omega\right]}=\det_+\left(-D_{\textrm{adj}}^{2}\right)\det_{-}^{N_f}\left(D^R_{\mu}\gamma_{\mu}+m\right).
\end{equation}
where $N_f$ is the number of (degenerate) flavors in the representation
$R$, and $\det_{\pm}$ indicates the imposition of periodic ($+$) or antiperiodic ($-$) boundary conditions in the Euclidean time direction. 
Calculation of the determinants gives the explicit formula
\begin{equation}
\Veff=-\frac2{\beta^4\pi^2}
\tr\Li_4(\Omega_{\textrm{adj}})
+N_f\frac{2m^2}{\beta^2\pi^2}\sum_{n=1}^{\infty}
\frac{(-1)^n}{n^2}K_2(nm\beta)\,\Re\tr\Omega_R^n,
\label{Veff}
\end{equation}
where 
$\Li_4(x)=\sum_jx^j/j^4$ and $K_n(x)$ is the modified Bessel function of the second kind.
$\Omega_{\textrm{adj}}$ and $\Omega_R$ are the matrices corresponding to \Eq{Omega} in the adjoint and $R$ representations.
One may easily express the diagonal values of $\Omega$,
$\Omega_{\textrm{adj}}$, and $\Omega_R$ in terms of the diagonal values of $\Phi=\textrm{diag}(\phi_1,\phi_2,\ldots)$ with $\phi_N=-\sum_{i=1}^{N-1}\phi_i$.
Choosing $R$ to be $S$, the two-index symmetric representation, the explicit formulas for the traces are
\begin{eqnarray}
L&=&\frac1N\tr\Omega=\frac1N\sum_{i}e^{i\phi_i}\\
\tr\Omega_S&=&\frac12\left[(\tr\Omega)^2+\tr\Omega^2\right]\nonumber\\
&=&\frac12\sum_{i\le j}e^{i(\phi_i+\phi_j)}\\
\tr\Omega_{\textrm{adj}}&=&|\tr\Omega|^2-1\nonumber\\
&=&\sum_{ij}e^{i(\phi_i-\phi_j)}-1
\end{eqnarray}
For $N=3$, where there are only two independent angles $(\phi_1,\phi_2)$, all of these may be expressed implicitly as functions of the real and imaginary parts of $L$, and plotted accordingly.

\begin{figure}[bt]
\includegraphics*[width=0.5\columnwidth]{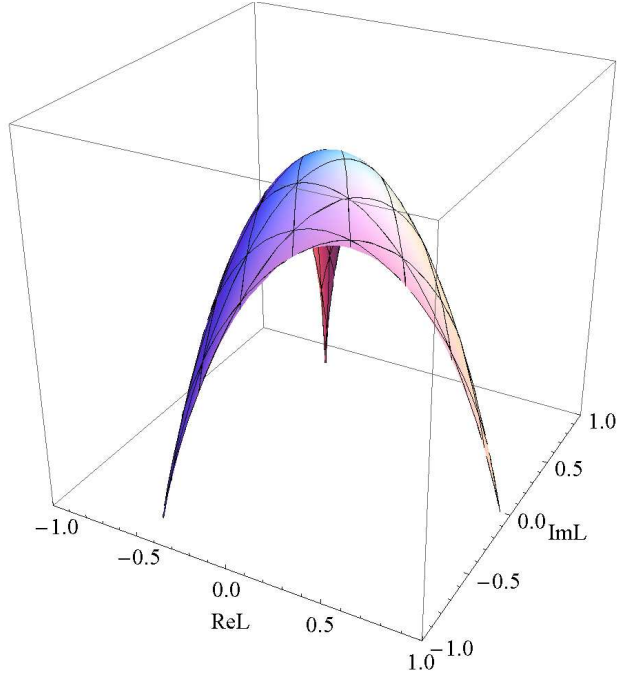}
\caption{One-loop effective potential $\Veff(L)$ for the
pure SU(3) gauge theory. \label{fig:Veff_pure}}
\end{figure}

\begin{figure}[ht]
\includegraphics*[width=.5\columnwidth]{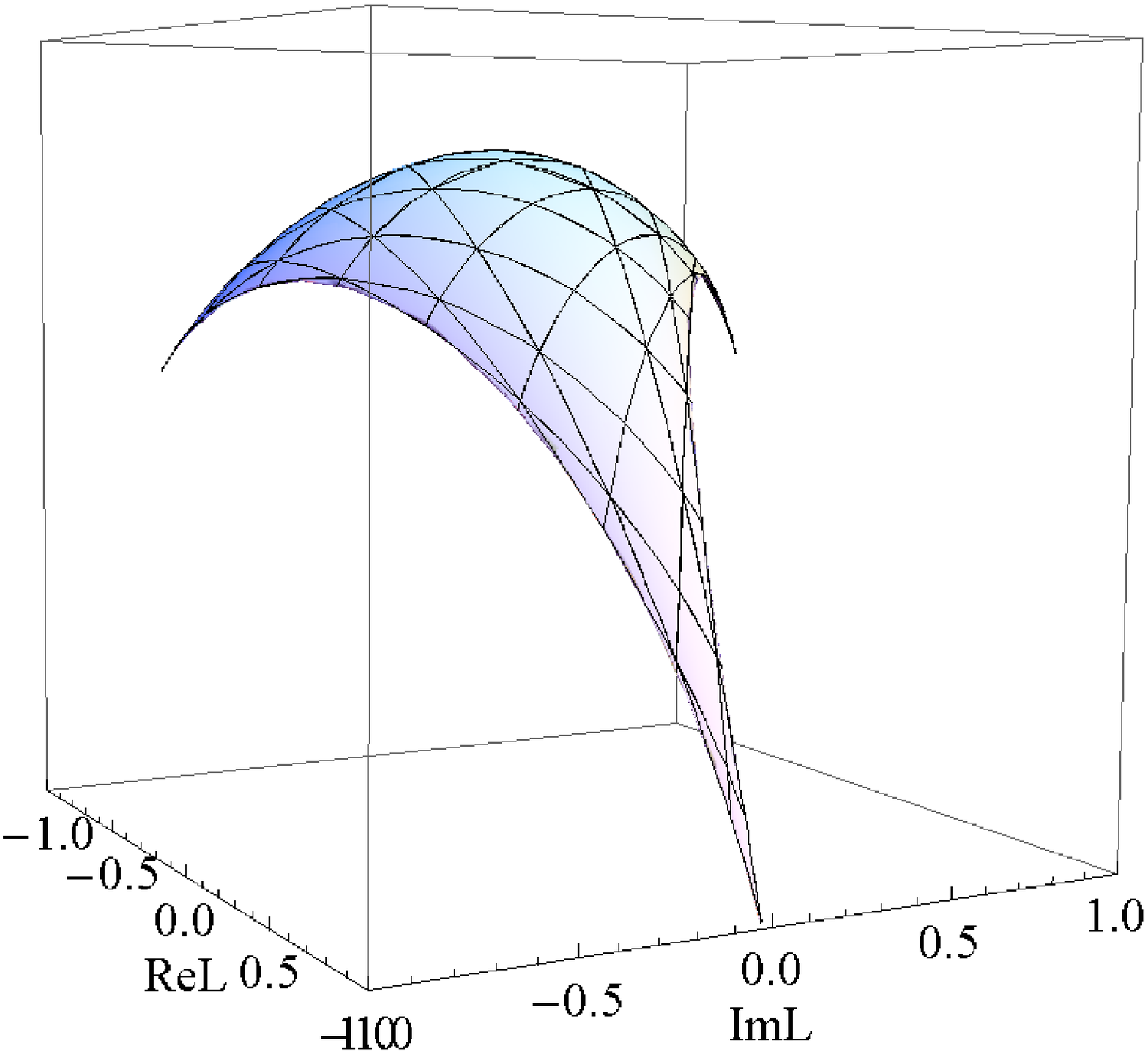}\\[10pt]
\includegraphics*[width=.5\columnwidth]{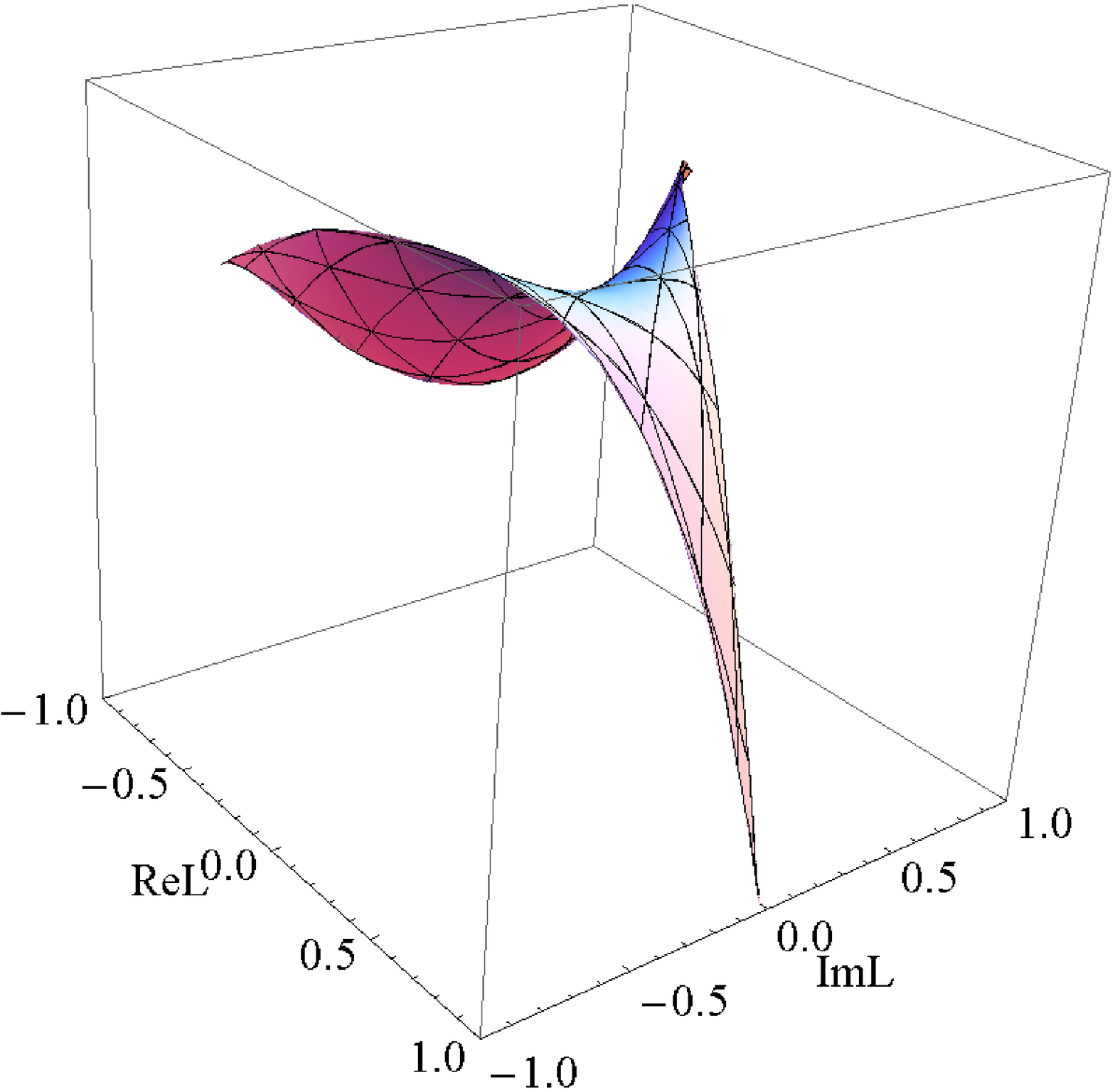}
\caption{One-loop effective potential for the SU(3)
gauge theory with very light fermions, $\beta m=0.002$. The fermions are in the
fundamental representation (top) and in the sextet
representation (bottom).\label{fig:Veff_ferm}}
\end{figure}

The first term in \Eq{Veff}, involving $\Omega_{\textrm{adj}}$, is the gluon contribution, invariant under
\ZN.
The second term, the fermion contribution, breaks \ZN~if $R$ has non-zero $N$-ality.
The fermion term decreases exponentially with increasing fermion mass.
To show its effect, we plot $\Veff$ for the pure SU(3) gauge theory and for
the theory with light triplet and sextet fermions, $\beta m=0.002$, in Figs.~\ref{fig:Veff_pure} and~\ref{fig:Veff_ferm}.
The fermion term lifts the degeneracy of the three Z(3) states, always favoring the $L=1$ state.
The other Z(3) states, $L=\exp(\pm2\pi i/3)$, become metastable;
in addition, the sextet case shows a new metastable state on the negative real axis of the $L$ plane.
As we will show, these metastable states survive into intermediate coupling.

%%%%%%%%%%%%%%%%%%%%%%%%%%%%%%%%%%%%%%%%%%%%%%%%%%%%%%%%%%%%%%%%%%%%%
\section{Large mass and strong coupling}
%%%%%%%%%%%%%%%%%%%%%%%%%%%%%%%%%%%%%%%%%%%%%%%%%%%%%%%%%%%%%%%%%%%%%

Another view of the effect of fermions on the effective potential for Wilson lines comes from the large-mass limit.
When the fermion mass is infinite, of course, the fermions decouple and the effective potential is \ZN-symmetric; its minimum is either a unique minimum at zero or a degenerate set of minima indicating spontaneous breaking of \ZN.
We can ask which way the effective potential tilts as the fermion mass is made finite \cite{Bartholomew:1983jv,Ogilvie:1983ss}.

We consider an SU($N$) lattice gauge theory with plaquette gauge action,
\begin{equation}
S_G=\frac{2N}{g^2}\sum_P\left(1-\frac1N\Re\tr U_P\right),
\end{equation}
and Wilson fermions in the representation $R$,
\begin{eqnarray}
S_F&=&\bar\psi M[U]\psi\nonumber\\
&=&\sum_n\bar\psi_n\psi_n-\kappa\sum_{n\mu}\bar\psi_n\left[(1-\gamma_\mu)U^R_{n\mu}\psi_{n+\hat\mu}
+(1+\gamma_\mu)U^{R\dag}_{n-\hat\mu,\mu}\psi_{n-\hat\mu}\right].
\end{eqnarray}
On integrating out the fermions, one obtains for the partition function
\begin{equation}
Z=\int dU\, e^{-S_G[U]}\,\Det\,M[U].
\end{equation}
When the fermions are massive one may expand in the hopping parameter
$\kappa=(2m+8)^{-1}$ to see their effect.
One obtains an effective action for the links that is a sum over closed paths (of length $\ell$) on the lattice,
\begin{equation}
\Seff[U]=S_G[U]+\sum_\ell\sum_{{\cal C}_\ell}\frac{\kappa^\ell}\ell
\tr\Gamma_{{\cal C}_\ell}\tr U^R_{{\cal C}_\ell}.
\label{hopping}
\end{equation}
Here $\Gamma_{{\cal C}_\ell}$ is a product of $(1\pm\gamma_\mu)$ around the path ${\cal C}_\ell$, and $U^R_{{\cal C}_\ell}$ is the product of link variables (in representation $R$) around the path.
A finite-temperature lattice has finite extent $N_t$ in the Euclidean time direction (and much larger in the spatial directions).
Thus there are two kinds of closed path in the sum in \Eq{hopping}.
Those that do not wind around the periodic time direction contribute terms that, like $S_G$, are invariant under \ZN.
Those that wind $k$ times contribute terms that transform as
\begin{equation}
S_{{\cal C}_\ell}\to z^{jk}S_{{\cal C}_\ell},
\end{equation}
where, again, $j$ is the $N$-ality of $R$.
This is the explicit breaking of \ZN~by the fermions.

Let us take for definiteness a lattice with $N_t=4$.
Then the first terms in the sum over paths in \Eq{hopping} are
the plaquette term and the real part of the Wilson line, both in the $R$ representation.
The latter has the form
\begin{equation}
\delta\Seff=-16N_f\kappa^4\sum_{\bf n}\Re\tr\Omega^R_{\bf n}
\equiv\sum_{\bf n}\Vf(L_{\bf n}),
\label{SeffStrong}
\end{equation}
where
\begin{equation}
\Omega^R_{\bf n}=\prod_{j=1}^{N_t}U^R_{({\bf n},j),\hat0}
\end{equation}
is the product of timelike links at spatial location {\bf n}.
The minus sign in \Eq{SeffStrong} comes from the antiperiodic boundary condition on the fermion field, which can be implemented by changing the sign of $(1\pm\gamma_0)$ on the last link of the lattice.

Considering again the sextet of SU(3), we 
note that when $L$ is real,
\begin{equation}
\Vf(L)\propto-\Tr \Omega_S=3L(1-3L).
\end{equation}
Near the origin, the fermions push $L$ in the negative direction.
Compare this to fermions in the fundamental representation, for which $\Vf(L)$ is simply proportional to $-\Re L$: It pushes $L$ in the positive direction.

We can illustrate this further by considering the strong-coupling limit, where we drop the plaquette term $S_G$; for simplicity, we also ignore the induced plaquette term that comes from the sum in \Eq{hopping}.
Then all that is left in the partition function,
\begin{equation}
Z=\int dU\, \exp \left[-\sum_{\bf n}\Vf(L_{\bf n})\right],
\end{equation}
is the measure and the local potential $\Vf(L)$.
The effective potential is then
\begin{equation}
\Veff=-\sum_{i<j}\ln[1-\cos(\phi_i-\phi_j)]+\Vf(L).
\end{equation}
Examination of $\Veff$ shows immediately that the minimum is near zero, displaced up or down the real axis as we have stated.

In the strong-coupling, confining phase the effective potential $\Veff(L)$ for the pure gauge theory has a unique minimum at zero.
The addition of the \ZN-symmetric terms in \Eq{hopping} to the effective action will move the boundary of the confining phase but within this phase the minimum remains at zero (by definition).
As we have seen, the addition of the \ZN-breaking term \Eq{SeffStrong} does move the minimum.
If the fermions are color triplets, the minimum is pushed up the real axis.
It is possible that as the lattice coupling is weakened the minimum moves continuously towards $L=1$, which is its weak-coupling destination.
In this case there is a continuous crossover between confining and non-confining regimes.
This is indeed what is seen for light quark masses, while for heavy quarks the transition is first-order.

Color sextet fermions, on the other hand, push the minimum of $\Veff$ in the negative direction. 
This makes it more likely that a barrier between this minimum and that at $L\simeq1$ persists in the transition region and makes the transition strongly first-order.
This is consistent with the results of Ref.~\cite{DeGrand:2008kx}, where a strong transition was found for all quark masses.

\section{Intermediate coupling: more phases?}

We have found two differences in the effective action $\Veff(L)$ created by triplet and sextet fermions.
In weak coupling, sextet fermions create an effective potential with a complex shape, admitting additional metastable states.
In strong coupling with large masses, the sextet fermions push $L$ in the negative real direction.
This raises the possibility that the condensate in the intermediate coupling region might condense in directions other than the positive real axis.
This indeed seemed to be the case in our earlier study~\cite{DeGrand:2008kx}.
In this section we will show results that indicate that while new
states may be reached in a quench from the disordered strong-coupling phase, these states are not equilibrium states. They can be made to decay to the true vacuum, which has a positive, real condensate $\svev{L}$.

Following the method of \cite{DeGrand:2008kx}, we simulated the
SU(3) gauge theory with $N_f=2$ flavors of sextet fermions.
The action is a plaquette gauge action with Wilson--clover fermions; our program, derived from the MILC code, uses the hybrid Monte Carlo algorithm.
We worked with $8^3\times4$ lattices with couplings chosen to frame the finite-temperature phase boundary: $5.25\leq\beta\leq5.6$ and
$0.145\leq\kappa\leq0.155$.
In Ref.~\cite{Shamir:2008pb} we found that $\kappa_c$ was between
0.164 and~0.169 in this range.  There we determined the tadpole coefficient $u_0$ self-consistently for each value of $\beta$ at $\kappa_c(\beta)$, reaching values between 0.881 and~0.889; here we set
$u_0=0.887$ throughout for simplicity.

\begin{figure}[bt]
\includegraphics*[width=.6\columnwidth]{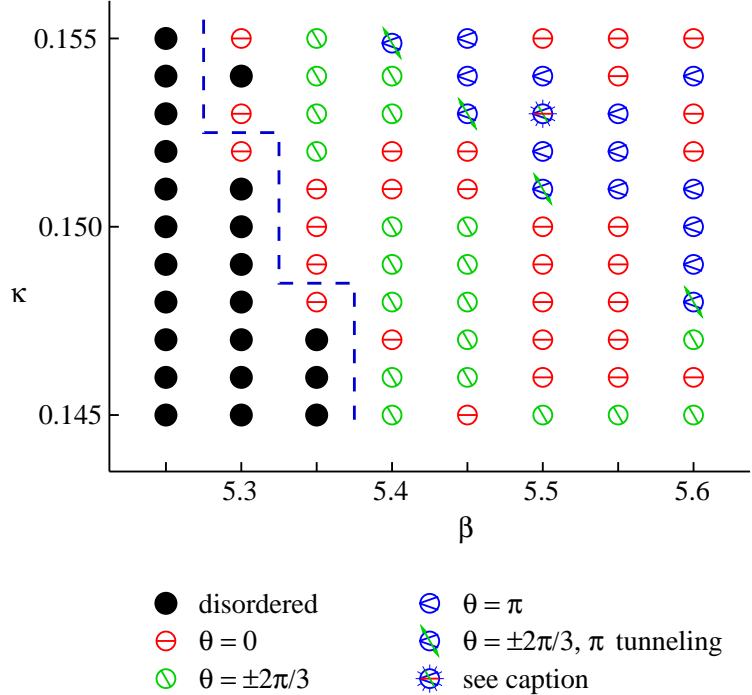}
\caption{Results of quenching from a disordered state.  Symbols indicate the orientation of the Wilson line condensate $\svev L$ following a quench from a disordered configuration.
The dashed line is the boundary between equilibrium phases---disordered vs.~$\theta\approx0$---determined with mixed starting configurations.
The last symbol in the legend marks a $(\beta,\kappa)$ value where two different states were reached in two runs: a state with tunneling between $\theta\approx\pm2\pi/3$ and $\theta\approx\pi$, and a state with $\theta\approx0$.
\label{fig:phases}}
\end{figure}

Quenching from a strong-coupling state creates an intricate ``phase diagram,'' shown in Fig.~\ref{fig:phases}.
There appear to be four weak-coupling states, distinguished by the direction that the condensate $\svev L$ takes in the complex plane: 
\begin{enumerate}
\item $\svev L$ real and positive (phase angle $\theta\approx0$);
\item $\svev L$ condensed in the direction
$\theta\approx+2\pi/3$;
\item $\svev L$ condensed in the direction
$\theta\approx-2\pi/3$;
\item $\svev L$ real and negative ($\theta\approx\pi$).
\end{enumerate}
The two states characterized by $\theta\approx\pm2\pi/3$ are related by charge conjugation $C$ and they exhibit spontaneous breaking of $C$.
They are approximately related to the $\theta\approx0$ state by the (explicitly broken) Z(3) symmetry, while the state with $\theta\approx\pi$ is not.
We find that $\svev L$ in the $\theta\approx\pi$ state is close in value to
$\Re\svev L$ in nearby $\theta\approx\pm2\pi/3$ states, which leads us to
conclude that the $\theta\approx\pi$ state is an Ising-like disordering of
the $\theta\approx\pm2\pi/3$ states that restores $C$ symmetry.

\begin{figure}[t]
\includegraphics*[width=.5\columnwidth]{b550_k147.eps}
\caption{Quench from disordered state to $\theta\approx0$ in the weak-coupling phase.
\label{fig:histories1}}
\end{figure}
\begin{figure}[t]
\includegraphics*[width=.5\columnwidth]{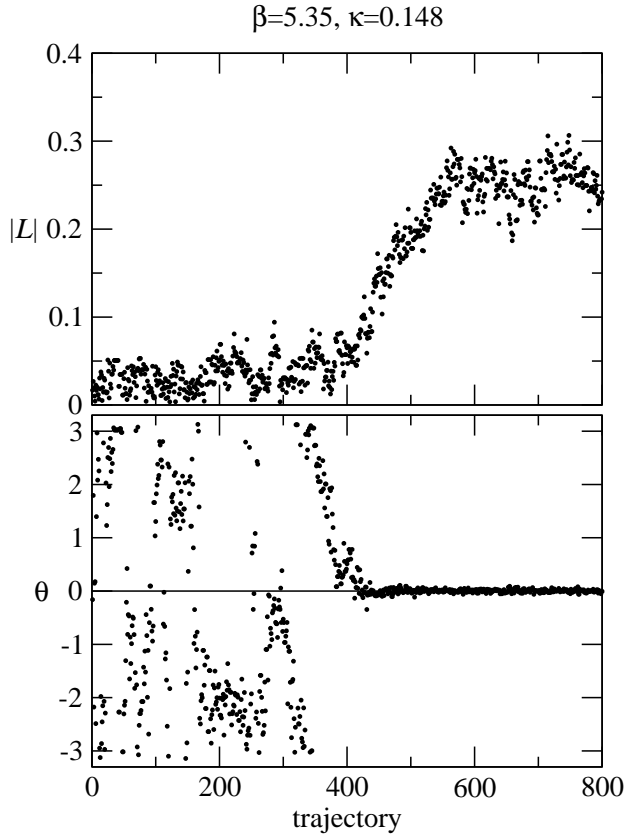}
\caption{Quench from disordered state to $\theta\approx0$ at stronger coupling than in Fig.~\ref{fig:histories1}.
In this case the true equilibrium state turns out to be disordered. The large value of $\svev{|L|}$ in the metastable state indicates that the phase boundary is first-order.
\label{fig:histories2}}
\end{figure}
\begin{figure}[t]
\includegraphics*[width=.5\columnwidth]{b560_k145.eps}
\caption{Quench from disordered state to $\theta\approx2\pi/3$ in the weak-coupling phase.
\label{fig:histories3}}
\end{figure}
\begin{figure}[bt]
\includegraphics*[width=.5\columnwidth]{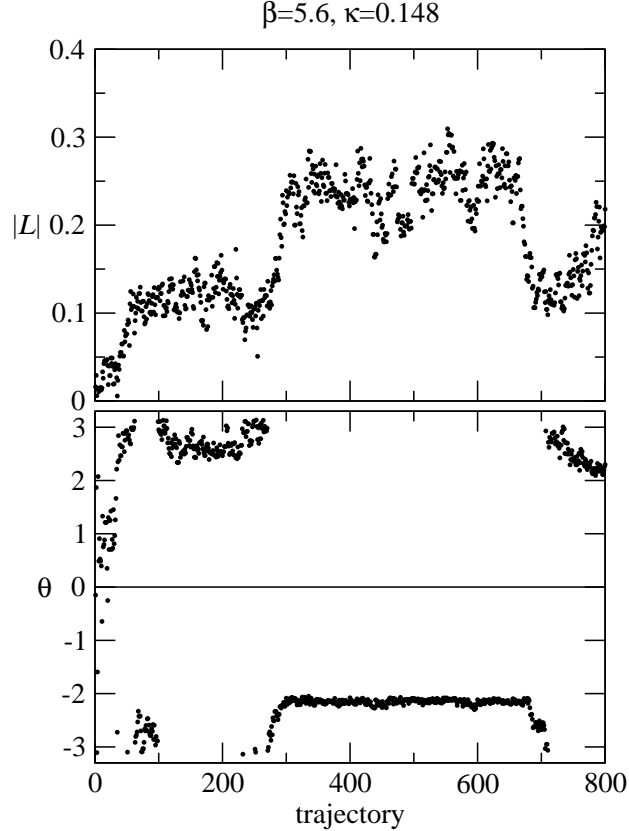}
\caption{Quench from disordered state to $\theta\approx\pm2\pi/3$ in the weak-coupling phase, showing tunneling between the two states
through $\theta\approx\pi$.
\label{fig:histories4}}
\end{figure}
\begin{figure}[t]
\includegraphics*[width=.5\columnwidth]{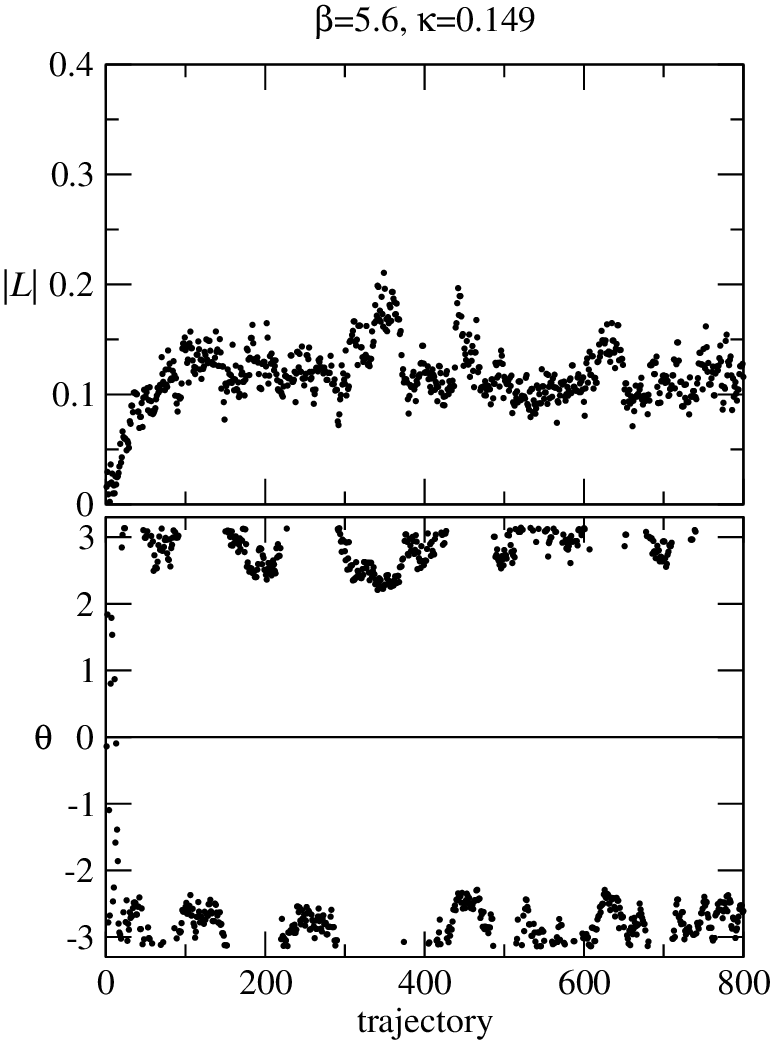}
\caption{Quench from disordered state to $\theta\approx\pi$ in the weak-coupling phase.
\label{fig:histories5}}
\end{figure}
We show in Figs.~\ref{fig:histories1}--\ref{fig:histories5} some histories of the quench to the various states.
Especially interesting are histories that show tunneling among the different weak-coupling states.
These are marked as coexistence points on the phase diagram.
%Tunneling between $\theta=2\pi/3$ and $\theta=\pi$ indicates that $\theta=\pi$ is a partially disordered state combining $\theta=2\pi/3$ with $\theta=-2\pi/3$; there seems to be an Ising-like transition restoring $C$ symmetry in the $\theta=\pi$ state.
We also found at least one point where different final states emerged in different runs from the same start.
(We did not search for this phenomenon systematically.)

It is quite curious that the phase diagram shows such a complex interleaving of the weak-coupling states.  
It is even more curious that the right-hand edge of the phase diagram exhibits all three states, even though weak-coupling analysis shows that only the $\theta=0$ state is stable in the limit.
These features raise the suspicion that we are dealing with metastable states.
A simple test proves that this is so.
Choosing a pair of adjacent couplings with different states indicated in the diagram, we can take a configuration from one coupling as an initial configuration for simulation at the other.
Almost any pair so chosen gives very large hysteresis, an almost complete failure to tunnel into the putative equilibrium state.

\begin{figure}[t]
\includegraphics*[width=.5\columnwidth]{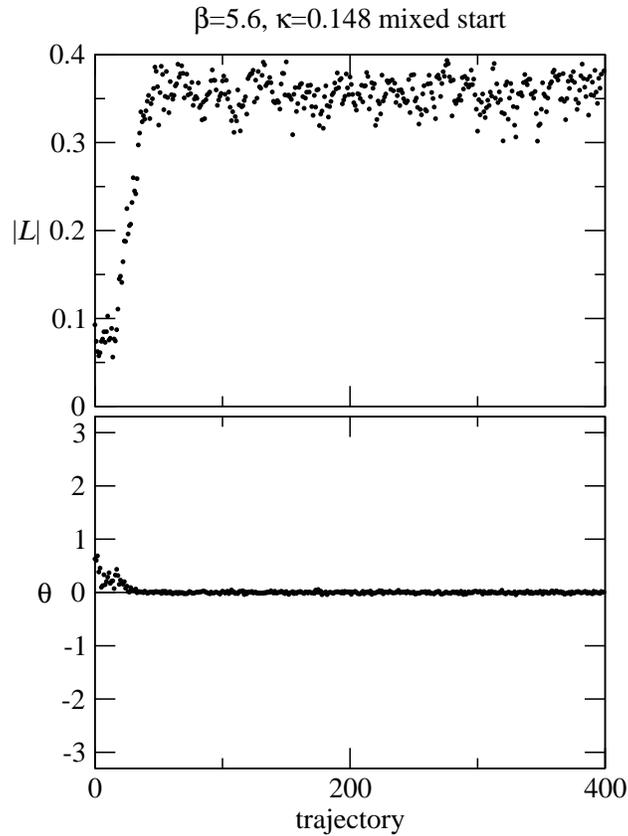}
\caption{Quench from mixed start in the weak-coupling phase.
\label{mixed}}
\end{figure}

We settled the issue by the simple expedient of using mixed starting configurations.
For each $(\beta,\kappa)$ where the quench resulted in $\theta\not\approx0$, we created a starting configuration by filling half the lattice with the field configuration coming from the quench and the other half with a $\theta\approx0$ configuration from a nearby coupling.%
\footnote{We also ran such checks on the $\theta\approx0$ states near the confinement phase boundary.}
The result was that {\em every\/} simulation in the ordered phase ended up in a $\theta=0$ state (see Fig.~\ref{mixed}).
This procedure also established with confidence the location of the true boundary between the disordered and the ordered phases, as shown in Fig.~\ref{fig:phases}.

\section{Summary}

Weak coupling analysis shows that the SU(3) gauge theory with sextet fermions possesses metastable ordered states with $\theta\approx\pm2\pi/3$, which break $C$ symmetry, and a state with $\theta\approx\pi$.
A hopping-parameter expansion shows that these fermions pull the Wilson line condensate $\svev L$ towards the negative real axis; thus the disordered phase tends towards $\theta=\pi$ as well.
It is thus not unreasonable to suppose that near the phase transition, the ordered phase might fall into an orientation other than $\theta\approx0$, possibly breaking $C$ symmetry.
Our numerical work displayed all these ordered states, but indicates that the only stable state (in the region we examined) is $\theta\approx0$, agreeing with the weak coupling limit.

The deflection of the disordered phase towards $\theta=\pi$ indicates that there is a barrier in the Landau effective potential separating this minimum from the weak-coupling minimum.
This argues for a first-order transition as the two phases compete, rather than the smooth crossover seen in QCD with light triplet quarks.

We emphasize that we have examined here the gauge theory at physical finite temperature, that is, with anti-periodic boundary conditions in Euclidean time for the fermions.
Our results should not be confused with those for periodic boundary conditions, that is, for two-index fermions in a finite box.
The latter indeed show an equilibrium condensate in directions other than $\theta\approx0$, with the possibility of spontaneous breaking of $C$ symmetry.

\section*{Acknowledgments}
We thank T.~DeGrand, J.~Myers, M.~Ogilvie, and Y.~Shamir for helpful
conversations.
This work was supported 
by the Israel Science Foundation under
grant no.~173/05. Our computer code is based on version
7 of the publicly available code of the MILC collaboration
\cite{MILC}.
%%%%%%%%%%%%%%%%%%%%%%%%%%%%%%%%%%%%%%%%%%%%%%%%%%%%%%%%%%%%%%%%%%%%%

%%%%%%%%%%%%%%%%%%%%%%%%%%%%%%%%%%%%%%%%%%%%%%%%%%%%%%%%%%%%%%%%%%%%%
\end{document}